\documentclass[epjST]{svjour}
\usepackage{graphicx}
\begin{document}

\title{
  Effect of grafting on the binding transition of two flexible polymers
}
%
\author{Johannes Zierenberg\thanks{\email{johannes.zierenberg@itp.uni-leipzig.de}} 
   \and Katharina Tholen\thanks{\email{katharina.tholen@itp.uni-leipzig.de}}
   \and Wolfhard Janke\thanks{\email{wolfhard.janke@itp.uni-leipzig.de}} }
\institute{Institut f\"ur Theoretische Physik, Universit\"at Leipzig, Postfach 100\,920, D-04009 Leipzig, Germany}

\abstract{
  We investigate the binding transition of two flexible polymers grafted to a
  steric surface with closeby end points. While free polymers show a
  discontinuous transition, grafting to a steric flat surface leads to a
  continuous binding transition. This is supported by results from Metropolis
  and parallel multicanonical simulations. A combination of canonical and
  microcanonical analyses reveals that the change in transition order can be
  understood in terms of the reduced translational entropy of the unbound
  high-temperature phase upon grafting.
} 
\maketitle
\section{Introduction}
\label{secIntroduction}
To study physical or chemical properties of polymers, one typically needs to
locally fix them. In the context of semiflexible polymer bundles, one can
experimentally rely on a multitude of methods such as fluorescence microscopy,
optical tweezers, atomic force microscopy, light scattering, electron
microscopy, and X-ray diffraction~\cite{schnauss2016}. For example, semiflexible
biopolymer bundles may be attached to polystyrene beads for a contact-free
manipulation via optical tweezers. Actin filaments are a good candidate to study
bundles of semiflexible polymers and their response to physical
forces~\cite{strehle2011,huber2012}. The stiffening of actin bundles can be
adjusted by cross-linking proteins~\cite{claessens2006,bathe2008}. Due to the
stiff nature of biopolymers, this can be well-described using the worm-like
chain model~\cite{heussinger2010}. This further allows one to study the binding
mechanism of directed polymers~\cite{kierfeld2003,kierfeld2005} or polymers
under tension~\cite{benetatos2014}, as well as force-induced desorption or
unzipping~\cite{kierfeld2006}.

Introducing attractive interactions among all monomers along the chain including
self-avoidance leads to so-called self-attractive semiflexible (theta)
polymers~\cite{zierenberg2016rev}. This formulation allows studies of the
competition between collapse and bundling, e.g., in the context of DNA
molecules~\cite{iwataki2004} where, depending on DNA and condensing agent
concentration, bundles of finite thickness were observed both numerically and
experimentally. In general, it was observed that bundles of semiflexible
polymers form twisted structures, which may be attributed to interaction-surface
maximization~\cite{heussinger2010,turner2003,grason2007,zierenberg2015epl}.  The
tendency to form polymer bundles requires sufficient (effective) stiffness; for
flexible polymers both static and dynamic reasons prevent bundle
formation~\cite{zierenberg2015epl,stevens1999}.

For the study of single or isolated polymer behavior, one often grafts the
chains to substrates (see, e.g., Refs.~\cite{rief1997,tress2013}). Depending on
the setup, an interaction with neighboring chains cannot be completely
excluded. In contrast, increasing the density of grafted polymer systems,
semiflexible (theta) polymers were shown to exhibit a rich phase space forming
toroidal, archway and tower micelles~\cite{bright1999,pham2005,benetatos2013}
for which they need to locally form bundles and which depends on both chain
length and density. A recent approach to systematically study the competition
between long-range charge repulsion and short-range attraction involves grafted
polymers that are allowed to move freely on the surface~\cite{benetatos2016},
finding finite-size bundles as well as formation of infinite bundles depending
on the range of repulsive interactions.

Here, we study the binding of two flexible polymers grafted closeby to a steric
surface. This is in contrast to the directed or semiflexible approaches to
study binding and allows one to compare to the uncorrelated motifs that originate
within aggregation of flexible polymers~\cite{zierenberg2015epl}. We notice that
there has been a somewhat equivalent use of the terms
aggregation~\cite{junghans2009,zierenberg2014jcp,mueller2015} and
binding~\cite{kierfeld2003,kierfeld2005}. Here, we refer to binding as the
process of two polymers attaching to each other, in our case even flexible
polymers. If specific inter-polymer interactions are considered this may quickly
lead to effects also characterized as zipping (see, e.g.,
Refs.~\cite{kumar1995,baiesi2001,leoni2003,ferrantini2011,walter2012,kumar2016}).
Even more interesting is the equivalence between two-polymer binding of directed
polymers and adsorption~\cite{kierfeld2003}. This should qualitatively remain
valid also for flexible polymers~\cite{zierenberg2017}, especially if one
imagines the crossover scenario of a flexible polymer adsorbed to a nanowire,
equivalent to the stiff limit of a polymer chain~\cite{vogel2015,gross2015}, or
a flexible polymer adsorbing to a flexible surface~\cite{karalus2011}. In fact,
it was shown that grafting alters the first-order-like adsorption transition
to a second-order-like transition~\cite{moddel2011}. We will show below that
the analogous scenario holds true for the binding of two flexible polymers,
where close-by-grafting may lead to single-polymer behavior.

The rest of the paper is organized as follows: After introducing the
investigated model and employed methods in Sec.~\ref{secMethods}, we present our
results in the canonical (Sec.~\ref{secCanonical}) and microcanonical
(Sec.~\ref{secMicrocanonical}) ensemble. We finish with our conclusions in
Sec.~\ref{secConclusions}.

\section{Model and Methods}
\label{secMethods}
A generic coarse-grained polymer model is a linear bead-spring homopolymer. In
principle, this allows one to model systems with linear chains from simple
synthetic polymers such as polyethylene to biological biopolymers such as DNA
or actin filaments if stiffness is included. In our case, monomers are connected
by the finitely extensible nonlinear elastic (FENE) potential, 
\begin{equation}
  V_{\rm FENE}(r)=-\frac{K}{2}R^2\ln\left(1-[(r-r_0)/R]^2\right),
  \label{eqModelFene}
\end{equation}
acting between bonded monomers. This is locally harmonic but diverges for
$|r-r_0|\rightarrow R$, where $r_0=0.7$ is the ``equilibrium'' bond length (if
no other potential was involved), $R=0.3$ and $K=40$. Non-bonded monomers
interact via the Lennard-Jones potential, 
\begin{equation}
  V_{\rm
  LJ}(r)=4\epsilon[(\sigma/r)^{12}-(\sigma/r)^6]=\epsilon[(r_0/r)^{12}-2(r_0/r)^{6}],
  \label{eqModelLJ}
\end{equation}
where $\sigma=2^{-1/6}r_0$ connects the non-bonded length scale to the
equilibrium bond length and $\epsilon=1$ sets the energy scale. The repulsive
$r^{-12}$ term models self-avoidance, while short-range attraction results from
the combination with the attractive $r^{-6}$ term. In total this mimics implicit
solvent. In accordance with the literature, we cutoff and shift the
Lennard-Jones potential to zero at $r_c=2.5\sigma$. This allows the usage of a
domain decomposition.
We focus on flexible polymers in this work, which undergo a collapse
transition at the theta temperature $T_{\theta}$. The total potential energy is
then obtained as the sum of interaction potentials. 

\begin{figure}
  \centering
  \includegraphics[width=0.3\textwidth]{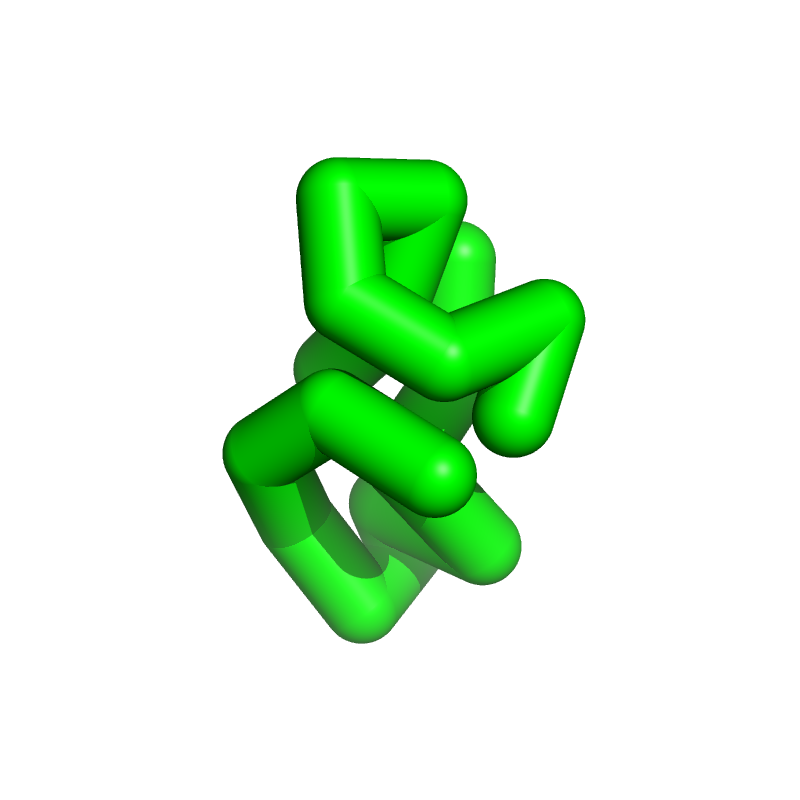}
  \hfill
  \includegraphics[width=0.3\textwidth]{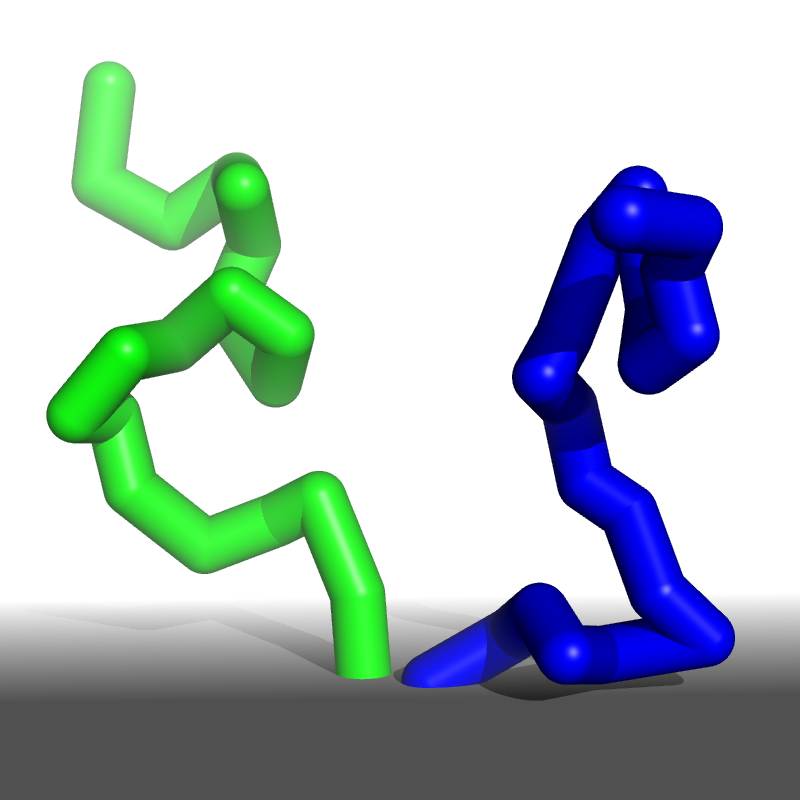}
  \hfill
  \includegraphics[width=0.3\textwidth]{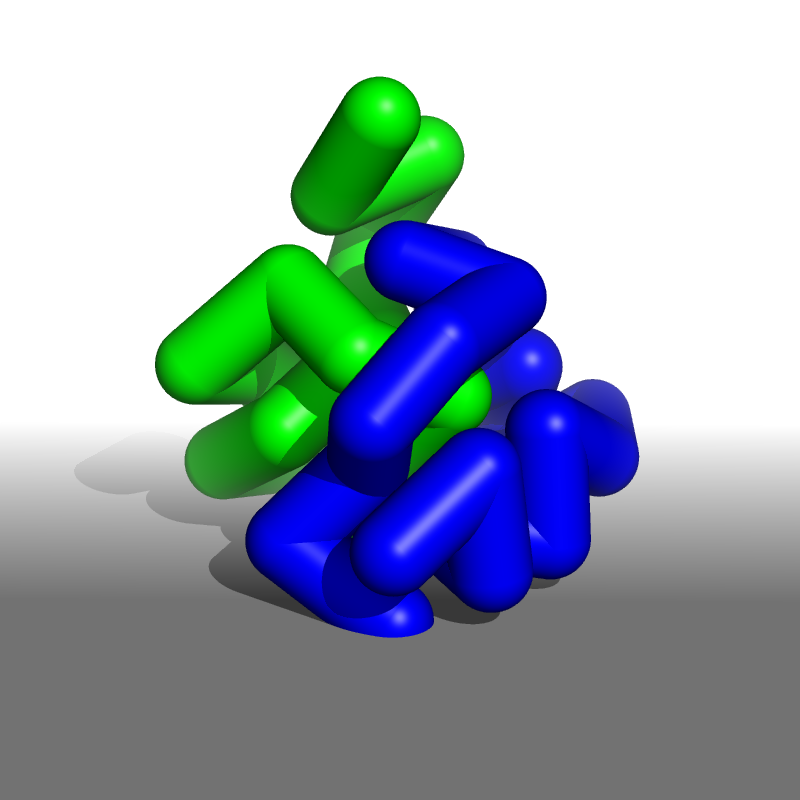}
  \caption{%
    Snapshots of a single polymer of length $N=20$ in the globular phase below
    the collapse transition (left, $T=0.7$) and two polymers of length $N=20$
    grafted to a steric surface above (middle, $T=4$) and below (right, $T=0.7$)
    the binding transition.
    \label{figSnapshots}
  }
\end{figure}

If the polymers are grafted, we fix one of their end points at a steric surface
covering the $x$-$y$ plane at $z=0$. No monomer is allowed to cross this
geometric constraint, i.e., $z_i<0$ is forbidden. No further interactions with
the (inert) surface are assumed. The end points are grafted with distance
$d=r_0$ and are immobilized. If the polymers are not grafted, we enclose them in
a cubic box of side length $L$ with steric walls. For an illustration see
Fig.~\ref{figSnapshots}.

We employ Markov chain Monte Carlo simulations to estimate thermodynamic and
structural properties in equilibrium. The phase space is commonly reduced to the
conformational phase space (or state space) by omitting the kinetic-energy
contributions. In principle, this reduction produces the same canonical
expectation values and thus the same canonical transition points. It will,
however, be relevant for the (conformational) microcanonical analysis (see
Sec.~\ref{secMicrocanonical})~\cite{zierenberg2016}. The first method we consider is the canonical
Metropolis algorithm~\cite{metropolis1953}, where proposed updates from a state
with potential energy $E_p$ to a state with potential energy $E_p^\prime$ are
accepted with the probability $\min(1,\exp[-\beta(E_p^\prime-E_p)])$, where
$\beta=1/k_{\rm B}T$ is the inverse temperature (we generally set $k_{\rm
B}=1$). A more elaborate multicanonical approach~\cite{bergMuca,jankeMuca} is to
consider a generalized ensemble, replacing the Boltzmann weight by a weight
function $W(E_p)$, which is iteratively adapted to yield a flat histogram
$H(E_p)$ for the acceptance probability $\min(1,W(E_p^\prime)/W(E_p))$. In this
formulation, the optimal result would correspond to the inverse conformational
density of states $W(E_p)\approx\Omega(E_p)^{-1}$. This can be seen if one
expresses the partition function in terms of a potential-energy integral, i.e., 
\begin{equation}
  Z_{\beta}=\int dE_p\Omega(E_p)e^{-\beta E_p} \rightarrow Z_{\rm muca} = \int dE_p\Omega(E_p)W(E_p).
\end{equation}
An estimate of the density of states in turn yields direct information about the
conformational microcanonical ensemble, since the microcanonical inverse
temperature is defined in terms of the density of states (see
below)~\cite{gross2001,janke1998,junghans2006,schnabel2011}.  Canonical
expectation values are obtained by histogram and time-series
reweighting~\cite{janke2012}. We follow the multicanonical approach with a
parallel implementation~\cite{zierenberg2013cpc}. For recent overviews of
generalized ensemble methods in the field of polymers see, e.g.,
Refs.~\cite{zierenberg2016rev,janke2016}.

Our Monte Carlo updates include short-range single-monomer shifts, long-range
polymer displacements (if applicable), and local bond rotations. If for any
monomer a shift below the surface ($z<0$) is proposed, the update is rejected by
the steric wall constraint. This generates an effective (entropic) repulsion
without any (energetic) potential. 
Error bars are obtained from statistical fluctuations taking into account the
integrated autocorrelation times (for the Metropolis data) and from Jackknife
error estimation (for the representative multicanonical data points obtained
from time-series reweighting)~\cite{janke2012}. From the multicanonical data we
additionally estimate high-resolution data points by histogram reweighting,
which suffers minor discretization effects from binning a continuous energy
domain. These form the connecting lines in the presented canonical results
below.

\section{Results}
\label{secResults}
We present our numerical results about the effect of end-point grafting on the
binding transition of two polymers from different viewpoints.  First, we start
with the canonical picture showing the result of grafting on the example of the
average energy and average end-to-end distance per polymer. Second, we change to
the microcanonical picture, which will allow us to draw qualitative conclusions
about the order of the finite-size binding transition.

\subsection{Canonical}
\label{secCanonical}
\begin{figure}
  \centering
  \includegraphics{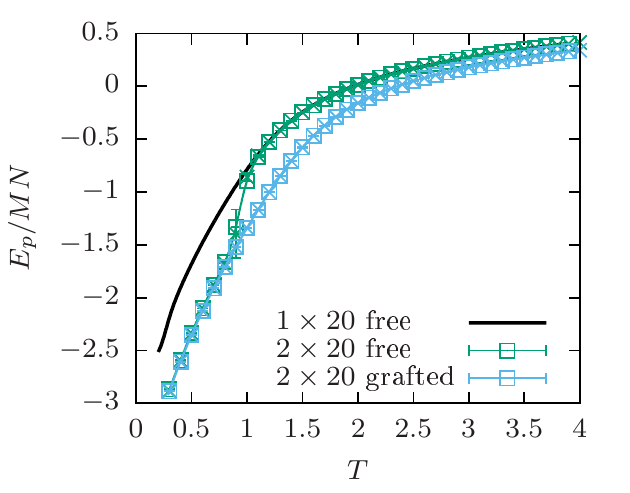}
  \includegraphics{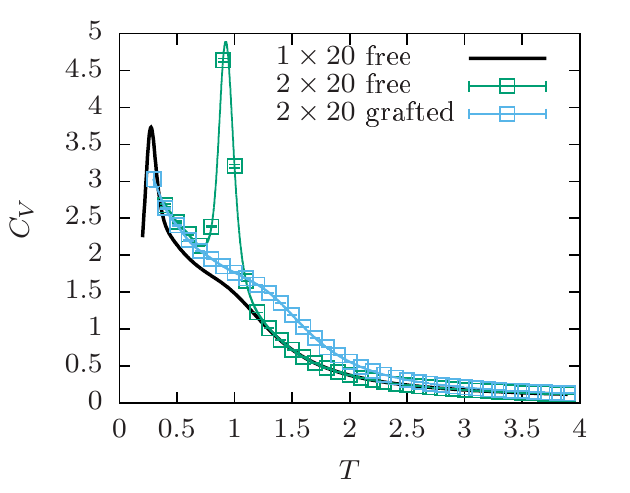}
  \caption{%
    Estimates of the canonical average potential energy and its derivative, the
    specific heat, for chains of length $N=20$ from parallel multicanonical
    (squares and lines) simulations. 
    For the energy we show in addition results from Metropolis simulations
    (crosses), which are in perfect agreement but suffer precision directly at
    first-order like transitions.
    The single-chain behavior is compared to the binding of two chains, both
    freely moving in a box ($\rho=10^{-4}$) and grafted to a steric surface.
    \label{figCanonicalEnergy}
  }
\end{figure}
In the first step we will focus on polymer systems with $M$ polymers of length
$N=20$. We compare the temperature curves of a single (isolated) free polymer,
two polymers at density $\rho=MN/L^3=10^{-4}$ in a cubic box with steric walls
(i.e., $L\approx74$),
and two polymers grafted at a steric surface at distance $d=r_0$, the bond
length defined in Eq.~(\ref{eqModelFene}).
Figure~\ref{figCanonicalEnergy} shows the average potential energy per monomer
$E_p/MN$ and its derivative, the specific heat $C_V=k_{\rm B}\beta^2(\langle
E_p^2\rangle-\langle E_p\rangle^2)/MN$. The single polymer case ($1\times20$
free) shows a
high-temperature extended regime signaled by high energies due to low number of
contacts. A temperature decrease causes the formation of monomer-monomer
contacts (compare Fig.~\ref{figSnapshots}~(left)), which leads to a decrease in
potential energy signaled by a shoulder in the specific heat, see
Fig.~\ref{figCanonicalEnergy}~(right) at $T\approx1$. This is the continuous
collapse transition. The peak in $C_V$ upon further temperature decrease
corresponds to a freezing transition into more compact states with even lower
potential energy. Adding a second polymer at a sufficiently low density (inside
a cubic box with steric walls, $2\times20$ free) reproduces the same
high-temperature behavior because the polymers may be considered isolated. The
qualitative behavior changes, initialized by a stronger energy decrease, at the
binding (or aggregation) transition. This is also signaled by a pronounced peak
in $C_V$ at $T=0.925(2)$, where the two polymers bind to each other and entangle to
maximize contacts and minimize energy. This is usually argued to be a
discontinuous, first-order transition due to the competition of entropy and
energy~\cite{junghans2009,zierenberg2014jcp,mueller2015,zierenberg2015epl}. Upon
further temperature decrease the polymer aggregate may be considered to behave
almost as a single polymer of length $N=40$ and a further freezing transition
may be expected. 

\begin{figure}
  \centering
  \includegraphics{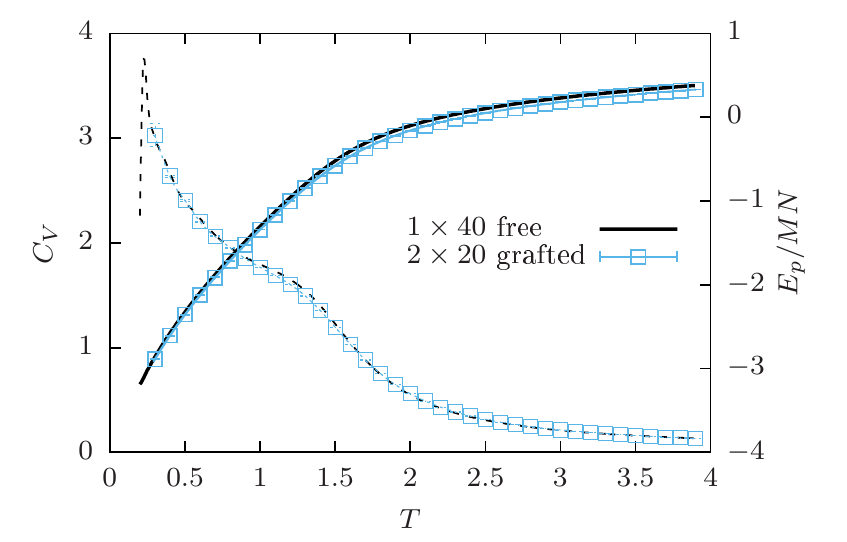}
  \caption{%
    Comparison of the canonical potential energy $E_p$ (solid) and its
    derivative $C_V$ (dashed) for a single polymer of length $N=40$ (black
    lines) and two grafted polymers of length $N=20$ (blue squares with error
    bars).
    \label{figCanonicalEnergy40}
  }
\end{figure}
If we now graft two end points at the steric surface ($2\times20$ grafted), we keep the qualitative
low-temperature behavior of two free polymers up to effective repulsive
interactions induced by the surface, which are barely noticeable in the average
energy. However, since the polymers are fixed to the surface they do not recover
the isolated single polymer behavior, having at least two monomers at the
surface in close contact. The discontinuous aspects in the binding transition of
two free polymers are now altered to features of a continuous binding transition
of two grafted polymers: We observe a continuous variation of the average energy
and a shoulder in the specific heat. Indeed, the shoulder in the specific heat
resembles the signal of a single polymer collapse.
Figure~\ref{figCanonicalEnergy40} shows canonical data for two grafted $N=20$
chains to be very similar to those of a single $N=40$ polymer. This can be
understood, considering that the two chains are grafted at a distance $d=r_0$,
equal to the bond length. Low-temperature conformations are consequently almost
indistinguishable, besides the reduction of symmetry by fixing a specific bond
to the outer shell of the globule. Even at high temperature the qualitative
behavior should be close to that of a single chain with effective repulsive
interactions due to a steric surface attached to its middle segment.  Thus, a
collapse-like binding mechanism is expected, in contrast to the entropy-energy
competition for the free counterpart.

\begin{figure}
  \centering
  \includegraphics{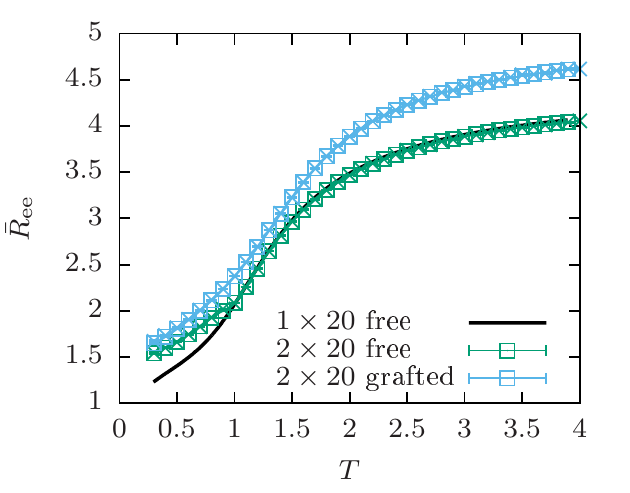}
  \includegraphics{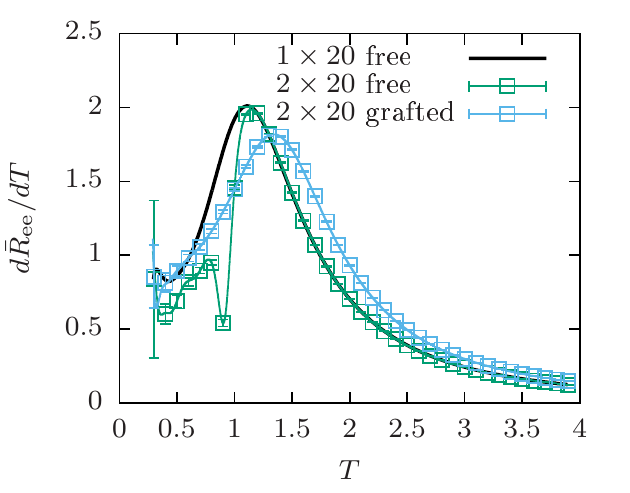}
  \caption{%
    Estimates of the canonical average end-to-end distance and its derivative for
    chains of length $N=20$ from parallel multicanonical (squares and lines)
    simulations. 
    For the end-to-end distance we show in addition results from Metropolis
    simulations (crosses), which are again in perfect agreement.
    The single-chain behavior is compared to the binding of two chains, both
    freely moving in a box ($\rho=10^{-4}$) and grafted to a steric surface (at
    distance $r_0$).
    \label{figCanonicalEnd}
  }
\end{figure}
This observation of a collapse-like process is supplemented in
Fig.~\ref{figCanonicalEnd} by looking at the average end-to-end distance,
defined as 
\begin{equation}
 \overline{R}_{\rm ee}=\frac{1}{M}\sum_{i=1}^M R_{\mathrm{ee},i},
\end{equation}
where $R_{\mathrm{ee},i}$ is the end-to-end distance of a single polymer, and
its thermal derivative $d \overline{R}_{\rm ee}/dT=k_{\rm B}\beta^2(\langle
\overline{R}_{\rm ee}E_p\rangle -\langle\overline{R}_{\rm ee}\rangle\langle
E_p\rangle)$. Again, we present data for a single polymer, two free polymers
(confined in a box) and two grafted polymers of length $N=20$. The
single-polymer collapse may be observed as a broad peak in $d \overline{R}_{\rm
ee}/dT$ at $T=1.111(6)$. The addition of a second polymer again shows a
high-temperature behavior consistent with the single polymer. The small
deviation is due to the confinement in a steric box. Simulations with periodic
boundary conditions for sufficiently dilute polymers show the identical
high-temperature behavior as for isolated
chains~\cite{zierenberg2014jcp,mueller2015}. At a density-dependent temperature
within the collapse peak, a sudden decrease to a local minimum signals the
binding (or aggregation) transition, here in the regime $T\in(0.9,1.1)$. This is
reflected also in the average end-to-end distance by a deviation from the
single-polymer behavior. The polymers are thus further extended in the aggregate
than in the globule. This signal is not as clear as for larger polymer numbers,
where a more drastic rearrangement occurs~\cite{zierenberg2014jcp}. 

When grafting the polymers, we can clearly see the effective repulsive
interaction of the steric surface in the average end-to-end distance, cf.\
Fig.~\ref{figCanonicalEnd}~(left). Grafting increases the average end-to-end
distance because elongations are only possible away from the surface and thus
away from the grafted end point. This is different for the free polymer, where
one end point may be in the center of a coil (low temperature) or where an
isotropic self-avoiding random walk would cross the surface (high temperature).
If one imagines the space of all (symmetric) conformations, then the steric
surface cuts off the half space. This removes the symmetric counterparts of the
extended chains but removes even more compact chains which form by crossing the
surface only with a subset of segments. Thus, the steric surface leads to a
stretching of a single chain.
In addition, grafting of two chains further analogously stretches the polymers
because of the local self-avoidance in the vicinity of the chain ends. The
average end-to-end distance of course shows differences to the single $N=40$
chain, being a local property which cannot be directly related. Instead, one can
image that the end-to-end distance of the free ends would show a similar
behavior, but this is outside of the present scope. The thermal derivative of
the average end-to-end distance shows again a broad transition peak around
$T=1.341(5)$. This further supports the coil-to-globule-like character of the
structural grafted-binding transition, which was conjectured above. 

%

\subsection{Microcanonical}
\label{secMicrocanonical}
\begin{figure}
  \centering
  \includegraphics{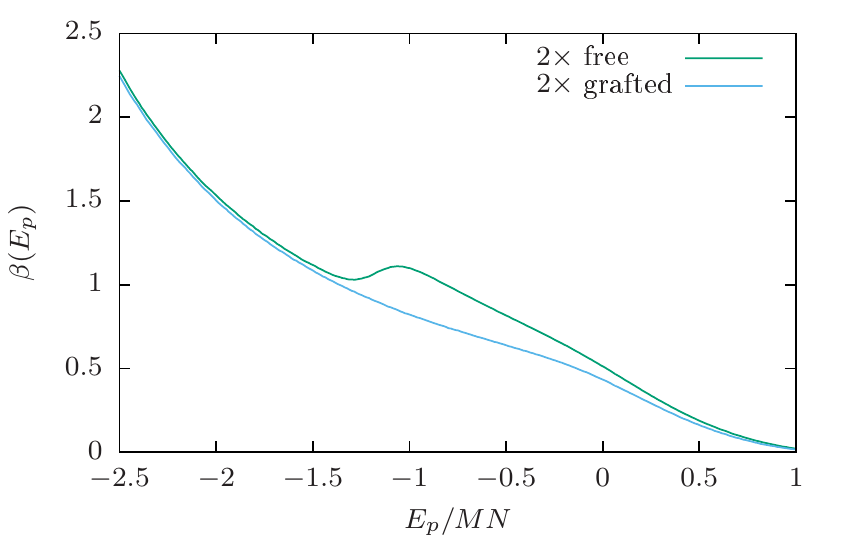}
  \caption{%
    Estimates of the microcanonical inverse temperature for two chains of length
    $N=20$ from parallel multicanonical simulations. The free chain shows the
    expected backbending typical for first-order finite-size transitions.
    Grafting the chains to a steric surface in close proximity (at distance
    $r_0$) diminishes the backbending and results in a second-order finite-size
    transition. 
    \label{figMicrocanBeta}
  }
\end{figure}
We move on to the microcanonical
viewpoint~\cite{gross2001,janke1998,junghans2006,schnabel2011}. Starting from a
conformational density of states it is useful to define the conformational
microcanonical entropy $S(E_p)=k_{\rm B}\ln\Omega(E_p)$. The derivative with respect
to the potential energy is $dS(E_p)/dE_p=k_{\rm B}\beta(E_p)$ with the microcanonical
inverse temperature, 
\begin{equation}
  \beta(E_p)=d\ln\Omega(E_p)/dE_p.
\end{equation}
This quantity encodes finite-size transitions in its inflection
points~\cite{gross2001,schnabel2011}. The example of two polymers of length
$N=20$ is shown in Fig.~\ref{figMicrocanBeta}.  For two free polymers at density
$\rho=10^{-4}$ we observe a back-bending in $\beta(E_p)$ at $E_p/MN\approx-1.2$,
which signals a first-order finite-size binding transition. This corresponds to
a convex intruder in $S(E_p)$ and can be translated into phase coexistence in
terms of a double-peaked canonical probability distribution
$P(E_p)\propto\Omega(E_p)e^{-\beta E_p}$~\cite{janke1998,schierz2016} with a
latent heat and a free-energy barrier. When grafting the polymers, the
back-bending vanishes, while the general behavior for low and high energies
remains compatible. We will see below that in the grafted case there is an
inflection point with negative slope, which suggests a second-order finite-size
binding transition.

%
\begin{figure}
  \centering
  \includegraphics{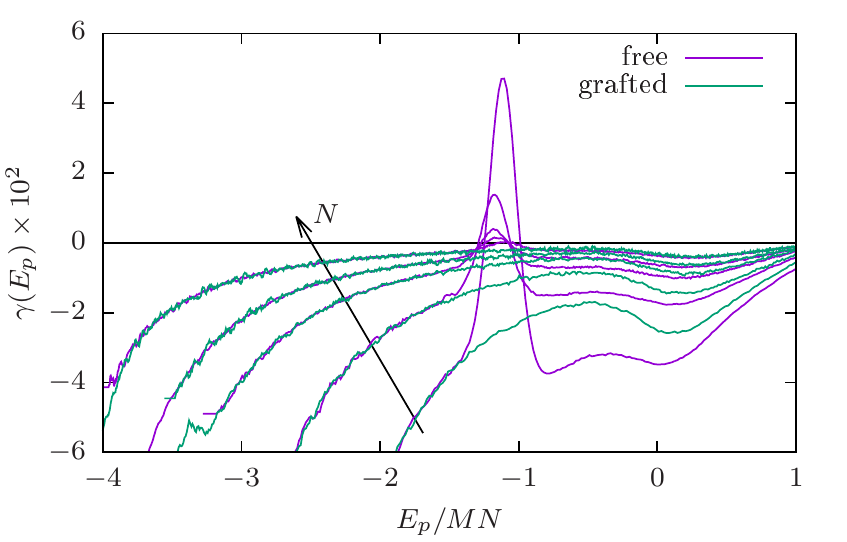}
  \caption{%
    Microcanonical inflection point analysis of the binding transition of two
    homopolymer chains for increasing length $N=\{13,20,30,40,60\}$. Compared
    are two chains in a steric box at fixed density $\rho=10^{-4}$ (free) such
    that the longest chain is still shorter than the linear box length, and two
    chains grafted to a steric surface at a distance $r_0$ (grafted). The free
    case shows positive peaks decreasing with polymer length, which signals a
    finite-size first-order transition possibly with a crossover to a
    second-order or a tricritical behavior (see discussion in text). The grafted
    case clearly shows negative peaks increasing with polymer length signaling a
    second-order finite-size transition, which may be expected to converge to
    the single-chain collapse in the limit of infinite chain length.
    \label{figMicrocanGamma}
  }
\end{figure}
The order of the finite-size transition may be well identified in the derivative
of the microcanonical inverse temperature,
\begin{equation}
  \gamma(E_p)=d\beta(E_p)/dE_p,
\end{equation}
where the inflection points correspond to peaks. A peak with a positive value
refers to a backbending in $\beta(E_p)$ and thus a first-order finite-size
transition, while a peak with a negative value signals a second-order
finite-size transition. In a proper thermodynamic limit, the peaks should
approach zero from above and from below, respectively.
Figure~\ref{figMicrocanGamma} shows $\gamma(E_p)$ for two polymers of various
chain length $N=\{13,20,30,40,60\}$; both either free in a cubic box with steric
walls at fixed density $\rho=10^{-4}$, or grafted to a steric surface at
distance $d=r_0$. We can clearly see that the finite-size binding transition of
two free polymers shows a positive peak for all chain lengths around
$E_p/MN\approx-1$ which rapidly decreases to zero. This peak corresponds to the
binding transition, where the small (negative) peak at higher energies may be
related to the single-polymer collapse transition. 

The binding peak of two free polymers actually seems to reach zero already for
finite system sizes. This is an indication that the considered thermodynamic
limit is not well defined. In fact, fixing the number of polymers $M$ and the
density $\rho=MN/L^3$ is not a suitable choice because the single-chain length
$N$ will inevitably exceed the linear size of the steric box, since then
$N\propto L^3$. As a consequence, the effective repulsive interaction induced
by the wall increases, leading to more prominent deformations of equilibrium
conformations. A more adequate approach towards infinite-size systems is the
limit of increasing polymer number $M$ at fixed polymer length
$N$~\cite{zierenberg2014jcp,zierenberg2016}.
%
In the present case, we want to focus on the finite-size binding transition of
$M=2$ polymers and merely illustrate the transition order for different chain
length, with $N<L$ for all cases (for $N=60$, $L\approx106$ still). 

Upon grafting, the situation changes and we only observe a single negative
maximum between $E_p/MN\approx-1$ and $E_p/MN\approx 0$. For small energies the
microcanonical behavior matches with the behavior for the free polymers because
the aggregate does not feel a density anymore. The single transition now
combines the collapse and binding transition due to the spatial vicinity and the
finite-size binding transition may be classified second order. In fact, the
systematic approach towards zero suggests that this formulation of increasing
chain length of two grafted polymers actually is a valid thermodynamic limit and
the binding transition of grafted polymers may be classified as a second-order
phase transition. We argued above that the binding transition is expected to
coincide with the single-polymer collapse for infinite chain length, consistent
with our results. General arguments imply also the same mean-field limit of
polymer solutions and isolated chains in the limit $N\rightarrow\infty$ (see,
e.g., Ref.~\cite{paul2007}), which was not accessible in the present scope and
is currently under investigation.

\section{Conclusions}
\label{secConclusions}
We have shown that grafting two polymers to a steric surface at a close
distance results in a second-order binding transition, which is in contrast to
the first-order like finite-size binding transition for free polymers in a
steric box. This is relevant for an experimental study of polymer binding,
where in vitro polymers would be commonly grafted. In this case, one will
neither observe a latent heat nor hysteresis effects associated with
first-order like transitions, which would be expected to occur in vivo. Still,
grafted polymers may be studied with respect to aggregate properties and their
dynamics which are expected to sufficiently coincide for observables which are
not directly influenced by the geometric constraint. One exception is the
average end-to-end distance and subsequently (no longer isotropic) geometric
properties.

Interesting effects may be anticipated for interacting surfaces, where the
binding of polymers would compete with the surface attraction. Connecting to
experimental setups, the surfaces may be considered both flat or curved, e.g.,
when grafting polymers to nanoparticles.

\acknowledgement{
The project was funded by the Deutsche Forschungsgemeinschaft (DFG) under Grant
No.\ JA~483/31-1. The authors gratefully acknowledge the computing time
provided by the John von Neumann Institute for Computing (NIC) on the
supercomputer JURECA at J\"ulich Supercomputing Centre (JSC) under Grant No.\
HLZ24.  
}


\end{document}